\documentstyle[twocolumn,aps,amssymb,epsf]{revtex}

\begin{document}
\draft

\title {Long Range Hydration Effects in Electrolytic Free
Suspended Black Films}

\author{D. Sentenac$^{1,2}$ and J. J. Benattar$^1$}

\address{$^1$Service de Physique de l'Etat Condens\'e,
Centre d'Etudes de Saclay,
91911, Gif--sur--Yvette, CEDEX, France\\
$^2$Present address:~F.O.M. Institute, Kruislaan 407, 1098 SJ
Amsterdam, The Netherlands}

\date{\today}

\maketitle

\begin{abstract}

The force law within free suspended black films made of negatively charged
Aerosol-OT (AOT) with added LiCl or CsCl is studied accurately using
X-ray reflectivity (ca. $1${\AA}). We find an electrolyte
concentration threshold above which a substantial additional repulsion
is detected in the LiCl films, up to distances of $100$ {\AA}. We
interpret this phenomenon as an augmentation of the Debye screening
length, due to the local screening of the condensed hydrophilic
counterions by the primary hydration shell. 

\end{abstract}


Because of their simple geometry and well defined double layered structure,
free-standing liquid films made of amphiphilic molecules are remarkable
model systems for colloidal science. As an example, the long range stability of
films made of neutral surfactants \cite{kocoex},
negatively charged surfactants \cite{dese}, negatively charged
amphiphilic polymers \cite{guscsemabe}, can be well accounted for by
simple mean-field electrostatic theories. A more complex situation
arises in black films (thicknesses in the range $60-300${\AA}) drawn
from sodium dodecyl sulfate (SDS), which exhibit, upon high addition
of sodium chloride \cite{exkokh}, or cesium chloride \cite{sede}, a
spectacular thinning transition when the distance between the
molecular planes, i.e., the separation, is $\approx 30$
{\AA}. Recently, progress has been made about the understanding of this
thinning transition, where it has  
been argued that the electrostatic correlations between the ions may
be predominant rather than attractive dispersion forces
\cite{sede,dehose}. This transition leads to the formation of the
Newton black film, where no liquid water remains in its core
\cite{bebe}. This kind of reversed membrane is stabilized by both
hydration and protrusion forces, the latter arising from molecular motion
of the surfaces \cite{books}. The nature and magnitude of these forces remain a
central question in colloidal science. Their investigation is an
extremely difficult task, as they both take place at the molecular
level. Studies reported on hydration forces, in systems where
protrusion forces do not occur, have revealed a short
range behaviour (they occur at separation $\lesssim 40$ {\AA}) and
a magnitude regulated by the number of hydration and the concentration
of the ions that bind to or condense electrostatically at the surfaces
(hard walls in this case)
\cite{books}. Complications arise when considering the ion size that
also influences the strength of surface forces. Size effects impose an upper limit 
on the density of the condensed counterions in the vicinity of a
charged surface. This can be accounted for in several ways: by the
introduction of an exclusion zone for the counterions, also known as
the Stern layer \cite{veov}, by the entropy of mixing \cite{boanho}, or by a
ion/ion hard core interaction \cite{kjma}. Whatever the method used, size effects
give rise to an additional repulsion, which is significant much beyond
the short range level, as they saturate the Debye screening of surfaces. 
Interestingly, ion size in water is related to its
hydration radius and therefore steric and hydration effects are
interdependent. However, as mentioned above, ion hydration
has been reported so far to give rise only to structural short range
interactions. We show in this Letter that ion hydration
lead also to long range behaviour. 

We measured accurately the long range force law within a free
suspended black film, drawn from an aqueous solutions of negatively charged
Aerosol-OT (AOT) (Sigma, purity $> 99\%$) with two kind of electrolyte added at high
concentrations ($0.05$, $0.1$, and $0.2$ M). The electrolytes used are
LiCl and CsCl (Fluka purity$> 99.5\%$). The two counterions 
differ  significantly in their number of hydration: $5-6$ water molecules per
Li$^+$ ion and $1-2$ per Cs$^+$ ion. However in water, Li$^+$ ions have an
hydrated  radius of $3.8$ {\AA} comparable to $3.3$ {\AA} for Cs$^+$
ions \cite{books}. All the measurements were performed with a
concentration of AOT at the Critical Micellar
Concentration (CMC), equal to $2.5$ mM. At the CMC,
micelles begin to form, and macroscopic films can be stabilized. In
all the systems studied, the added electrolyte
concentration is higher than of the surfactant by several orders of
magnitude. The contribution of the AOT self counterion Na$^+$ in the
interactions can therefore be neglected. 
Similar to \cite{wi} (after several
stages of purification), no minimum in the surface tension isotherm of
the pure AOT solution was detected.  The salts were annealed  at $500^o$C  before
use to remove any surface impurities. The water used comes from a  
milli-Q system. Its surface tension is $72.5$ mN/m at $22\pm
1^o$C. All the measurement were carried out at $22\pm 1^o$C. 

The force law between the molecular planes follows from the
measurements of the disjoining pressure and the equilibrium thickness
of the film using a very recent technique.
It combines the original porous plate method \cite{myjo,exsc}, with the
accurate determination of the thickness, using X-ray reflectivity
\cite{scbeko}. With the introduction of the X-rays, this apparatus
allows the investigation of forces at the molecular level
\cite{sescnebe}. The porous plate method uses a porous glass disk as a
liquid reservoir for the free-standing film. A hole drilled
through the disk is used as an holder for an horizontal liquid
film. The ensemble is enclosed in an airtight Plexiglas cell
partially filled with the solution to maintain a saturated vapor atmosphere. A
pressure  $\Delta P$ can be  applied on the film by
means of a syringe connected to the cell, whereas the
porous plate is connected to a capillary tube at the atmospheric
pressure  $P_o$ (Fig. 1). The pressure in the cell is $\Delta P + P_o$ and
$\Delta P$ is measured by means of a differential manometer with an
accuracy of $0.1$ millibar (mb). The disjoining pressure $\Pi_d$
of the film is the pressure difference between the
film reservoir and the gaseous phase. It is thermodynamically
related to the derivative of the Gibbs free energy of the film
\cite{der}. Experimentally, it is given by the relation $\Pi_d =
\Delta P - P_h$, where $P_h =  \Delta\rho gh - 2\sigma/r$ is the
hydrostatic pressure, $\sigma$ is the surface tension of the solution, $r$ is the
capillary tube radius, $g$ is the gravitational acceleration, and $\Delta\rho$
the density difference between the solution and the air.  The
hydrostatic pressure is of the order of $1$ mb and can be neglected in the present
study. The maximum pressure allowed is fixed by the Laplace pressure
of the  pores inside  the disk. We used porous plates (Robu) with mean
pore  radius of $0.75\,\mu$m allowing pressures $\lesssim 600$ mb. To
achieve an X-ray reflectivity
experiment in association with disjoining pressure measurements, the
films are formed with a modified version of the traditional porous
plate. Two hermetically sealed Kapton windows allow the passage of the
beam through the cell. The hole size of the disk is macroscopic
($15 \times 4$ mm$^2$) as imposed by the reflection of X-rays at grazing
angles. In addition, the film must stand on the extreme upper
surface of the disk holder. For that purpose the hole profile has
an inclination angle of $30^o$. To avoid any residual shadowing
effects due to the meniscus of the film, a slit of
$1\mu$m depth was added in the continuation of the hole, in the
direction of the beam. 

The cell is placed on the head of a four-circle diffractometer
in a horizontal geometry. A conventional
fine-focus copper tube is used as an X-ray source. The monochromator
 is a Si(111) crystal which selects the $CuK_{\alpha 1}$ line
($\lambda=1.5405$ {\AA}). The beam is collimated in the scattering plane
by an incoming slit ($100\mu$m) placed at $40$ cm from the source,
providing  a low divergence ($0.15$ mrad). A vertical
slit ($3$ mm) is used to limit  the width of the illuminated area on
the film. The reflected beam is detected by a scintillation counter
placed behind an analysis slit ($250 \mu$m) at a distance $40$ cm from
the center of the diffractometer. The incident and reflected beams
pass through vacuum flight paths.  
\begin{table}
\begin{tabular}{|c|c|c|c|}\hline
Region & Reduced density  & Thickness& Roughness\\
       &($\delta$$\times 10^{6}$) & {\AA}& {\AA}\\ \hline 
AOT layer& $4.2\pm 0.1$ & $11\pm 0.5$ & $3.3 \pm 0.5$\\ 
Aqueous core& 3.68 & $22 \pm 0.5$& $3.3 \pm 0.5$\\ \hline
\end{tabular}
\vskip 0.5 truecm
\caption{Structural parameters of a symetrical $3$-layer model, as
determined by X-ray reflectivity, of a free-standing black film drawn
from a solution of AOT ($2.5$ mM) with added $0.4$ M LiCl. The total
thickness is $44\pm 1$ {\AA}.}
\label{tabaot}
\end{table}

\begin{figure}
\centerline{\epsfxsize=200pt\epsffile{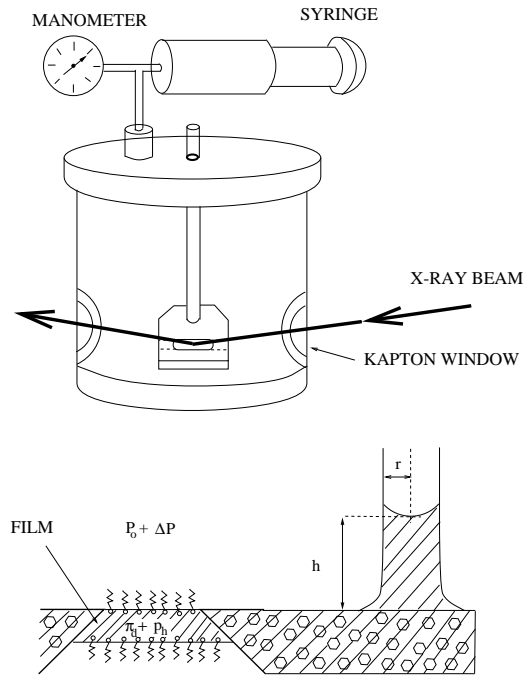}}
\caption{Scheme of the apparatus installed in the center of the
diffractometer. A macroscopic flat film ($15\times 4$ mm$^2$) is drawn at
the extreme upper surface of the hole of the porous plate initially
filled with the solution. The disjoining pressure of the film is
measured by the application of an additional pressure in the cell. The
film structure is determined by X-ray reflectivity.}
\label{cellaot}
\end{figure}
The specular reflectivity exhibits
Kiessig fringes resulting from the interference due to
large density gradients through the film. A film mosaic arises due to the
nonflatness of the holder which broadens the specular peak in the
transverse direction by about $20${\%}. 
Finally, the level of accuracy
achieved  with this apparatus is $\lesssim 1$ {\AA} for the
determination of the real thickness of the film. 
The analysis of the reflectivity profiles is performed
rigorously through the use of an optical formalism valid at all angles
\cite{born}, convoluted with the experimental Gaussian resolution,
including the film mosaic.
The film is described by a succession of homogeneous slabs in the
direction perpendicular to the film, each of them characterized by
three  parameters: thickness, electron density, and interfacial
roughness. The electron density is related to the refractive index via
the relation $\rho= 2\pi\delta~/~\lambda^2 r_e$, where $r_e$ is the
classic radius of electron, and $\delta$ the so-called reduced density.

The knowledge of the generic structure of AOT black films is the first
step for the $\Pi_d$-isotherms determination. No structure differences
has been detected whatever the electrolyte added and the thickness of the
aqueous core (the separation) is the only adjustable parameter during
pressure measurements.
Table 1 shows the parameters corresponding to the best fit
of the reflectivity curve 
\begin{figure}
\centerline{\epsfxsize=300pt\epsffile{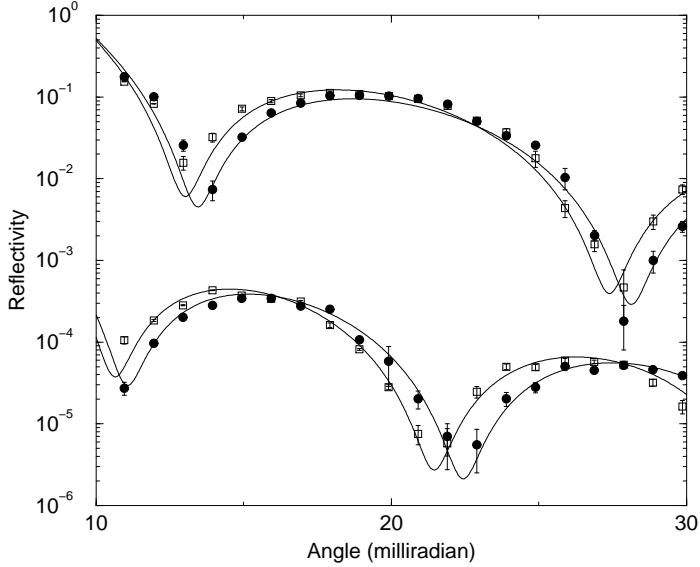}}
\caption{Successive X-ray reflectivity curves on a black film of AOT
($2.5$mM) with added $0.2$M LiCl; lower curves: $\Delta P = 200$mb (squares),
$\Delta P = 250$mb (circles). Upper curves (shifted): $\Delta P =
450$mb (squares), $\Delta P = 500$mb (circles). The theoretical fits
are in solid lines and indicate a thinning transition from  
$72\pm 1${\AA} to $69\pm 1${\AA} (lower curves) and from $61.5\pm
0.5${\AA} to $60\pm 0.5${\AA} (upper curves).}
\label{coupap}
\end{figure}
\begin{figure}
\centerline{\epsfxsize=300pt\epsffile{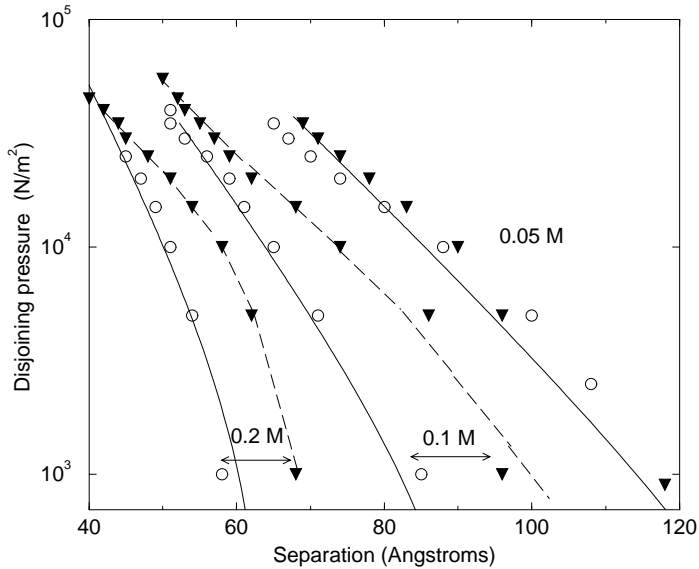}}
\caption{$\Pi_d$-isotherms of free-standing films  AOT ($2.5$ mM),
with from the right to the left, a concentration of
$0.05$, $0.1$, $0.2$ M of (circles) CsCl, and (triangles) LiCl;
(solid lines)  fits using Lipshitz theory for dispersion forces [21] and
Poisson-Boltzmann equation for electrostatic forces [22];
(dashed lines) guides to the eye.}
\label{isoaot1}
\end{figure}
recorded on a $44$ {\AA} thick $2.5$ mM AOT
black film with added $0.4$ M LiCl. The film is described by a
symetrical $3$-layer model distinguishing the aqueous core from the
two amphiphilic layers.  
The interfacial roughnesses are essentially due to the
thermal  fluctuations of the molecular layers \cite{bebe}.  The
contribution of the two surfactant planes to the overall thickness is
$22\pm 1${\AA}. 

The pressure was increased by steps of $50$ mb (until film rupture) every
$30$ mn to insure the film is in equilibrium. The films remain entirely
homogeneous over their whole area and react instantaneously to any 
pressure variation. Several successive scans on the LiCl systems are shown in
Fig.2. Note the high contrast of the Kiessig fringes which allow
the detection of thickness changes of $\lesssim 1${\AA}. The
$\Pi_d$-isotherms are plotted with respect to the
separation between the AOT molecular planes on Fig.3.  
One can first notice that the slope of all the $\Pi_d$-isotherms above
$0.05$ M of salt tends to bend at the lowest pressures ($\lesssim
10^4$ N/m$^2$). Such an effect has been previously observed on
$\Pi_d$-isotherms of SDS films with added CsCl  \cite{sede}, and is
the  signature
of the attractive dispersion forces. At $0.05$ M the $\Pi_d$-isotherms
with LiCl or CsCl are identical. At $0.1$ M of electrolyte
added, an aditionnal repulsion arises on the LiCl $\Pi_d$-isotherms up
to  separations of $100$ {\AA}. A similar gap is present at $0.2$ M of
electrolyte, keeping approximatively the same value of $\approx 10$ 
{\AA} up to $100$ mb.
As a result, the formation of the gap is both electrolyte and concentration
dependent. The fact that the film thickness depends on the electrolyte has already
been pointed out and related to the strength of the
binding of the counterions in the vicinity of the surface
\cite{jomysc,br}. Here, the
analysis of the disjoining pressure isotherms allows us to specify the
origin of this effect. 

The general shape of all the $\Pi_d$-isotherms with Cs$^+$
\begin{figure}
\centerline{\epsfxsize=250pt\epsffile{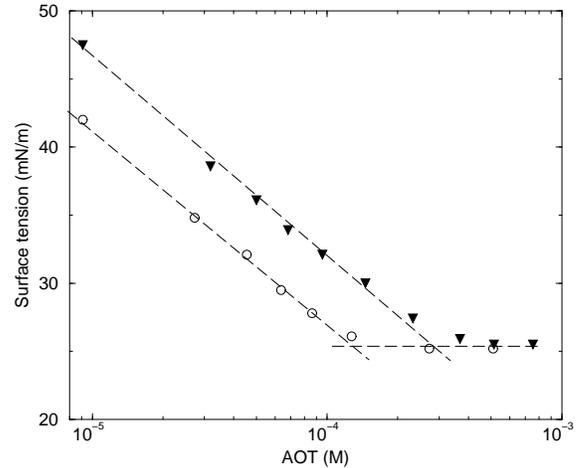}}
\caption{Surface tension isotherms, measured with the Wilhelmy
balance system, of AOT with added $0.2$ M of (circles) CsCl, 
(triangles) LiCl. The CMCs are located at the intersection between the
dashed lines.}
\label{isoaot}
\end{figure}
\noindent ions presents a rather good agreement with mean-field
electrostatic theory, neglecting size effects \cite{dorivr,chpawh}. The surface
potentials fitted  at the aqueous core/surfactant interface are in the range
$95-100$mV.  These potentials are similar to those predicted
in \cite{sede} for SDS films. However, above $0.1$ M the 
$\Pi_d$-isotherms obtained with Li$^+$ ions cannot be accounted for by these
theories. Adding a Stern layer in the soap head region of thickness
$4$ {\AA} (much larger than one would expect relative to a CsCl
electrolyte) does increase the electrostatic repulsion but not
sufficiently to explain the gap. Moreover, steric effects may
already be present in the films containing $0.05$ M of electrolyte,
which are several orders of magnitude above the CMC.  Therefore, the important
discrepancy observed cannot be simply due to a finite size effect.  

In fact, the gap that arises between the CsCl and LiCl $\Pi_d$-isotherms
can be directly related to the behaviour of the
corresponding surface tension isotherms. On Fig.4 are reported two
isotherms of  surface tension of  AOT solution containing $0.2$ M of
LiCl or CsCl. A clear gap can be seen between the two surface tension
isotherms indicating a higher adsorption of AOT molecules at the air/solution
interface, with Cs$^+$ ions present rather than Li$^+$ ions. In
addition, the CMC appears higher in the case of LiCL ($0.32$ mM) than CsCL 
($0.15$ mM). Much below the CMC, finite size effects are not supposed
to intervene. Therefore, the adsorption of AOT molecules will be
hindered by the hydration shell of the Li$^+$ counterions that appears
to weaken the electrostatic screening of the SO$_3^-$ groups. Thus, a basic
approach to explain the gap between the $\Pi_d$-isotherms would be an
augmentation of the Debye screening length of the LiCl systems by reducing the
net charge of the Li$^+$ counterions, induced by the local screening of
the surrounding water molecules. On another hand, one should also
evaluate the dehydration energy cost of the Li$^+$ ions in the screening
process of the SO$^-_3$ groups. An interesting point is that
at pressures $\gtrsim 10^4$ N/m$^2$ the gap between the
$\Pi_d$-isotherms diminishes and the slope of the Li$^+$
$\Pi_d$-isotherms becomes lower than for Cs$^+$, which is consistent
with an augmentation of the Debye screening length. Nevertheless, in
this part of the $\Pi_d$-isotherms, and at these high ionic strengths,
attractive ionic fluctuations may be also considered \cite{sede}. 

In conclusion, The deviations between the CsCl $\Pi_d$-isotherms and
LiCL $\Pi_d$-isotherms are the signature of a hydration
effect.  We have shown that it is significant up to distances of
$\approx 100$ {\AA}. Therefore, it appears that ion hydration
phenomenon gives rise, not only to structural short range forces, but also,
and similarly to finite size effects, has a long range behaviour. Its
action may be understood within the framework of the
electrostatic theory, by considering the local screening of the
hydrophilic ions by the surrounding water molecules.

We are very grateful to D. S. Dean, J. A. Hodges and M. Nedyalkov for
extremely  helpful discussions. We thank R. Tourbot for technical assistance.


\begin{references}

\bibitem{kocoex}{T. Kolarov, R. Cohen, D. Exerowa, Colloids Surf.
, {\bf 42}, No.1-2, 49 (1989).}
\bibitem{dese}{D. S. Dean and D. Sentenac, Europhys. Lett., {\bf 38}(9),
645 (1997).}
\bibitem{guscsemabe} {P. Guenoun, A. Schalchli, D. Sentenac, J. W. Mays, 
J. J. Benattar,  Phys. Rev. Lett., {\bf 74}(18), 3628 (1995).}
\bibitem{exkokh}{D. Exerowa, T. Kolarov and Khr. Khristov, 
Colloids Surf., {\bf 22}, 171 (1987).}
\bibitem{sede}{D. Sentenac and D. S. Dean, J. Colloid Interface Sci.,
{\bf 196}(1), 35 (1997).}
\bibitem{dehose}{ D. S. Dean, R. R. Horgan, and D. Sentenac,
J. Stat. Phys., {\bf 90} 3/4, 899 (1998).}
\bibitem{bebe}{O. B\'elorgey and J.J. Benattar, Phys. Rev. Lett., {\bf 66},
313 (1991).}
\bibitem{books}{J. Israelachvili, {\em Intermolecular and Surface Forces},
 (Academic Press, 1992).}
\bibitem{veov}{E. J. W. Verwey and J. Th. G. Overbeek, {Theory of
Stability of Lyophobic Colloids}, (Elsevier, Amsterdam, 1948).}
\bibitem{boanho} {I. Borukhov, D. Andelman, and H. Horland,
Phys. Rev. Lett., {\bf 79}(3), 435 (1997).}
\bibitem{kjma}{R. Kjellander S. Marcelja, J. Phys. Chem., {\bf 90}(7),
1230 (1986).}
\bibitem{wi} {E. F. Williams, J. Colloid Sci., {\bf 12}, 452 (1957).}
\bibitem{myjo} {K. J. Mysels and M. N. Jones , Discuss. Faraday Soc.,
{\bf 42}, 42 (1966).}
\bibitem{exsc} {D. Exerowa and A. Scheludko, C.R. Acad. Bulg. Sci.,
{\bf 24}, 47 (1971).}
\bibitem{scbeko} {Schalchli A., J. J. Benattar, T. Kolarov,
C.R. Acad. Sci. ser II, {\bf 319}, 745 (1994).}
\bibitem{sescnebe}{D. Sentenac, A. Schalchli, M. Nedylakov and J.J.
Benattar, Faraday Discuss., {\bf 104}, 345 (1996).}
\bibitem{der}{B. V. Derjaguin, N. V. Churaev, and V. M. Muller, 
{\em Surface Forces}, (Consul. Bureau, New York, 1987).}
\bibitem{born} {M. Born and E. Wolf, {\em Principles of Optics},
(Pergamon,  London, 1980).}
\bibitem{jomysc}{M. N. Jones, K. Mysels and P. C. Scholten,
Trans. Faraday Soc., {\bf 62}, 1336 (1966).}
\bibitem{br}{R. Krustev, D. Platikanov, M. Nedyalkov, Colloids
Surf. A, 123-124, 383 (1997).}
\bibitem{dorivr}{W.A.B. Donners, J. B. Rijnbout and A. Vrij, 
J. Colloid Interface Sci., {\bf 60}(3), 540 (1977).}
\bibitem{chpawh}{ D. Y. C. Chan, R. M. Pashley and L. R. White, 
J. Colloid Interface Sci., {\bf 77}(1), 283 (1980).}
















\end{references}
\end{document}